\documentclass[11pt,a4paper]{article}

\usepackage[a4paper, total={6in, 8in}]{geometry}
\usepackage{authblk}

\usepackage{float}
\usepackage{graphicx} 
\usepackage{hyperref}
\usepackage[cmex10]{amsmath}
\usepackage{%
	amsfonts,%
	amsmath,%
	amssymb,%
	amsthm,%
	bbm,%
	cite,%
	float,%
	etoolbox,%
	mathrsfs,%
	mathtools,%
	multirow,%
	pgf,%
	pgfplots,%
	tikz,%
}

\usepackage[normalem]{ulem}

\usepackage{booktabs}
\usepackage{algorithm}
\usepackage{algpseudocode}
\usepackage{bm}
\usepackage{tikz}
\usepackage{amssymb}

\usepackage{graphicx}
\usepackage{float}
\usepackage[caption=false,font=footnotesize]{subfig}

\title{
Boundary-Aware Stabilizer Scheduling for Distributed Quantum Error Correction
}

\author[$*$]{Sanidhya Gupta}
\author[$*$]{Sanidhay Bhambay\thanks{sanidhay21@gmail.com}}
\author[$\dag$]{Narges Alavisamani}
\author[$*$]{Neil Walton}
\author[$*$]{Thirupathaiah Vasantam}
\affil[$*$]{Durham University, UK}
\affil[$\dag$]{Georgia Tech, USA}
\date{}

\begin{document}
\maketitle
\begin{abstract}


It is anticipated that future quantum architectures will be modular, with quantum processors connecting multiple quantum processing units  (QPUs) via photonic interconnects. This creates heterogeneity between local and distributed entanglement, making error correction even more critical for reliable computation. In topological quantum error correction, such as color codes, the distributed setting creates \emph{seam} boundaries where parity checks require remote CNOT operations using heralded Bell pairs. These non-local checks are slower and noisier than bulk local checks because entanglement generation is probabilistic, causing data qubits to accumulate idle noise 
while waiting for remote operations. A natural way to reduce this overhead is to skip some seam measurements; however, doing so makes seam syndrome information stale and can degrade decoding. Thus, the central scheduling problem is to determine how frequently seam checks should be measured so as to balance remote-operation and waiting noise against syndrome staleness.

To address this trade-off, we develop a scheduling module that integrates directly into standard syndrome-extraction circuits. Within this framework, we consider two policies: Skip-Seam-$\tau$ (SS-$\tau$), which measures all bulk checks every round while measuring seam checks once every $\tau$ rounds and copying syndromes in skipped rounds from recent measurements, and Adaptive Skip-$\tau$ (AST), which selects the seam-refresh interval $\tau$ adaptively as a function of code distance and entanglement generation rate (EGR). We evaluate these policies on triangular color codes under circuit-level noise in Stim, explicitly including idling errors induced by Bell-pair generation delays.

Our simulations show that SS-$\tau$ and AST reduce the remote-operation overhead associated with non-local CNOTs and can lower the logical error rate (LER) relative to the Measure-All (MA) baseline, which measures all stabilizers in every syndrome-extraction round. For physical error rate $p=10^{-3}$, we identify an EGR regime in which both SS-$\tau$ and AST exhibit behavior consistent with fault-tolerant scaling, with LER decreasing as code distance increases. Across these regimes, SS-$\tau$ and AST outperform MA, indicating that appropriate seam-measurement scheduling can substantially narrow the performance gap between distributed and monolithic architectures.

\end{abstract}

 \section{Introduction}
Quantum error correction (QEC)~\cite{shor1995scheme,steane1996error,terhal2015quantum,nielsen2010quantum,bhambay2026adaptive} protects quantum information by encoding a logical qubit across many physical data qubits, while ancillary parity qubits are used to extract syndrome information and detect errors. 
Each QEC code defines a set of \emph{stabilizer generators}, whose measurement outcomes indicate whether local errors have occurred or not. Each stabilizer generator is measured using an associated ancilla qubit that interacts with the relevant data qubits through a short sequence of CNOT gates.
These gates map the joint parity of the data qubits onto the ancilla, which is then measured to produce a classical bit known as the \emph{syndrome} or parity.
Repeating this process over successive cycles yields a stream of syndromes that a classical decoder uses to infer and correct physical errors.
Stabilizer measurement is therefore a fundamental primitive in fault-tolerant quantum computation (FTQC): it dominates timing, communication, and control flow in each round of QEC~\cite{gottesman2009introduction,preskill1998reliable,aharonov1997fault,bombin2006topological,fowler2011two,landahl2011fault, kitaev2003fault,fowler2012surface}. 
As a result, the scheduling, latency, and reliability of stabilizer measurements directly affect the throughput and fidelity of fault-tolerant execution.

FTQC is typically analyzed in a monolithic architecture\cite{gottesman1997stabilizer,google2023suppressing}, where a single quantum processing unit (QPU) contains all data and ancilla qubits. 
In this setting, syndrome extraction proceeds through a homogeneous set of stabilizer measurements implemented using local gates and synchronized rounds.
In a monolithic setup, all parity checks have similar costs, and decoders expect a steady and regular stream of syndromes, so the default policy of measuring every check every round is both natural and efficient~\cite{chamberland2023techniques,smith2023local,alavisamani2024promatch,ravi2023better,das2022lilliput}.

Current QPUs, however, have only a few hundred physical qubits. 
To fully unlock the potential of quantum computing, we need QPUs with thousands, and eventually far more, high-quality qubits capable of executing millions of fault-tolerant operations~\cite{komoto2025quantum,smith-goodson2025ibm}.
A promising way to achieve this is through modular quantum architectures~\cite{figueiredo2021engineering,evered2023high}, in which a large quantum processor is constructed from smaller QPUs connected by quantum communication channels\cite{cirac1999distributed,van2013blueprint}. 
In such architectures, practical constraints naturally lead to distributed quantum error correction (DQEC)~\cite{sutcliffe2025distributed,chandra2025distributed,singh2024modular}, in which a single code patch is distributed across multiple QPUs connected by photonic interconnects. 
This creates two classes of stabilizer checks: {\em bulk} checks, which are local to a single QPU, and {\em seam} checks, which span QPU boundaries and require non-local CNOTs, typically implemented using shared Bell pairs,
as illustrated in Fig.~\ref{fig:remoteCNOT}. 
This architectural shift introduces an inherent asymmetry. 
Seam checks rely on Bell pairs that are both noisy and generated probabilistically.
As a result, non-local operations incur stochastic waiting times determined by the entanglement generation rate (EGR), during which data qubits may remain idle and accumulate decoherence, we call such noise as idle noise.
Consequently, seam checks are generally slower and less reliable than bulk checks.


This asymmetry has direct consequences for syndrome extraction. At low EGR, delays in Bell-pair generation lead to substantial idle-time decoherence on the data qubits participating in remote stabilizer measurements~\cite{cirac1999distributed}. In addition, seam checks are subject to the larger error rates of remote gates and readout. These effects jointly reduce the fidelity of seam-parity measurements and, when operations are synchronized, can also increase the latency of the entire syndrome-extraction round.

As a result, measuring seam stabilizers in every round can be costly and, in some regimes, counterproductive. This motivates considering schedules that skip seam measurements, thereby spreading the cost of distributed stabilizer operations over multiple rounds. Of course, this introduces a trade-off: while skipping reduces measurement-induced noise, it increases the risk of stale syndrome information, potentially allowing errors to accumulate undetected.

Skipping seam stabilizer measurements can be advantageous because, in the proposed Skip-Seam-$\tau$ (SS-$\tau$) scheme defined later in this section, the idle-noise cost associated with Bell-pair generation is incurred only once every $\tau$ rounds. Consequently, the average per-round idle error decreases approximately as $1/\tau$. By contrast, in the conventional Measure-All (MA) scheme, corresponding to $\tau=1$, seam stabilizers are measured in every round, so this waiting-induced idle-noise penalty is paid every cycle. Thus, in the low-EGR regime, increasing $\tau$ can reduce accumulated idle noise and improve logical performance. However, skipping seam measurements also makes boundary syndrome information stale, increasing the risk that parity flips go undetected for several rounds.



This trade-off leads to the central scheduling question in distributed QEC: \textit{how often should seam stabilizers be measured?} Bulk stabilizers are local and comparatively inexpensive, whereas seam stabilizers require non-local CNOTs and therefore incur additional gate, readout, and delay-induced errors. A good scheduling policy must therefore measure seam checks often enough to keep boundary information fresh, but not so often that remote-operation and waiting costs dominate the error budget.

More broadly, DQEC introduces a fundamental asymmetry between bulk and seam stabilizers, so stabilizer measurement can no longer follow a uniform measure-all policy. Instead, the frequency of seam-check measurements should be treated as a controllable design parameter if distributed architectures are to approach the performance of monolithic systems.

To address the problem of syndrome-measurement scheduling for DQEC, we introduce two seam-scheduling schemes that reduce the frequency of seam-check measurements:
\begin{itemize}
    \item \textbf{Skip-Seam-$\tau$} (SS-$\tau$): In this scheme, all bulk stabilizers are measured in every round. However, seam stabilizers are measured only once every $\tau$ rounds and are skipped for the remaining $\tau-1$ rounds. When a seam stabilizer is skipped, its ancilla is not entangled in that round, and the previous syndrome bit is copied forward to keep the detector model deterministic.

    \item \textbf{Adaptive Skip-$\tau$} (AST): In this scheme, all bulk stabilizers are again measured in every round, while the seam-refresh interval is chosen adaptively as a function of the code distance $d$ and the EGR. In the low-EGR regime, seam stabilizers are refreshed less frequently to reduce remote-operation overhead, whereas in the high-EGR regime they are refreshed more often, since entanglement generation is no longer the dominant bottleneck.
\end{itemize}
Both scheduling policies keep the code and decoder unchanged; they only modify how frequently seam information is sampled. We compare these two policies against the traditional \textit{Measure-All} (MA) scheme, in which both bulk and seam checks are measured in every round and which serves as the baseline for evaluation.

\subsection{Main contributions}

The central contribution of this work is to show that seam-measurement scheduling is a meaningful design parameter in DQEC; the SS-$\tau$ and AST policies serve as concrete examples that establish the utility of this idea. To our knowledge, this is the first work to systematically formulate and evaluate stabilizer-measurement scheduling for DQEC. 

We perform all simulations in Stim~\cite{gidney2021stim} for triangular-patch color codes under both local and remote noise, explicitly including idling errors arising from Bell-pair generation delays across QPUs. Within this setting, we introduce a lightweight scheduling controller that determines, in each round, which stabilizers to measure and which to skip, while integrating directly into Stim circuits without modifying the underlying code or decoder. We then propose two seam-scheduling policies: SS-$\tau$, which measures all bulk stabilizers every round while refreshing seam stabilizers once every $\tau$ rounds, and AST, which chooses the seam-refresh interval adaptively as a function of code distance and EGR. Across the simulated code distances, both SS-$\tau$ and AST achieve lower LER than MA in the low-to-moderate EGR regime by reducing the number of noisy remote seam operations. Furthermore, for physical error rate $p=10^{-3}$, we identify an EGR regime in which both policies exhibit behavior consistent with fault-tolerant error correction over the simulated distances.

Our results show that periodic and adaptive seam-scheduling policies reduce the non-local CNOT load and can significantly improve the LER over a range of EGRs at a fixed $p$.

\subsection{Related works}
Large-scale FTQC typically assume monolithic devices with uniform gate timing and synchronized rounds \cite{gottesman1997stabilizer,google2023suppressing}. To overcome the limited qubit counts of current monolithic architectures, many works propose modular or distributed quantum architectures in which multiple QPUs are linked by photonic interconnects~\cite{cirac1999distributed,van2013blueprint,figueiredo2021engineering,evered2023high}. Within modular architectures, DQEC places a single code patch across several modules and performs seam (inter-QPU) stabilizers using non-local operations. Recent works explore protocols, thresholds, and resource trade-offs for DQEC. For example, a distributed Floquet code is explored in~\cite{sutcliffe2025distributed}, a color code in~\cite{chandra2025distributed} and for the surface code\cite{singh2024modular}. 
In all these works, the main idea is to measure all stabilizers, both bulk and seam, and scheduling of stabilizers is not the primary focus of these works.

\subsection{Organization}

The remainder of the paper is organized as follows:
Section~\ref{sec:bg_goal} reviews the background and problem setting.
Section~\ref{sec:dqec_frmwrk} presents the distributed QEC model and the proposed seam-scheduling framework.
Section~\ref{sec:color_code_part} describes how the triangular color code is partitioned across multiple QPUs.
Section~\ref{sec:methodology} introduces the simulation methodology.
Section~\ref{sec:num_rslts} discusses the simulation results, and
Section~\ref{sec:conclusion} concludes the paper.

\section{Background and Goal}
\label{sec:bg_goal}
In this section, we discuss the required background knowledge before we present our model in the next section.
\subsection{Color codes}
\begin{figure}[h!]
\centering
  \medskip
  \centering
  \includegraphics[width=7cm]{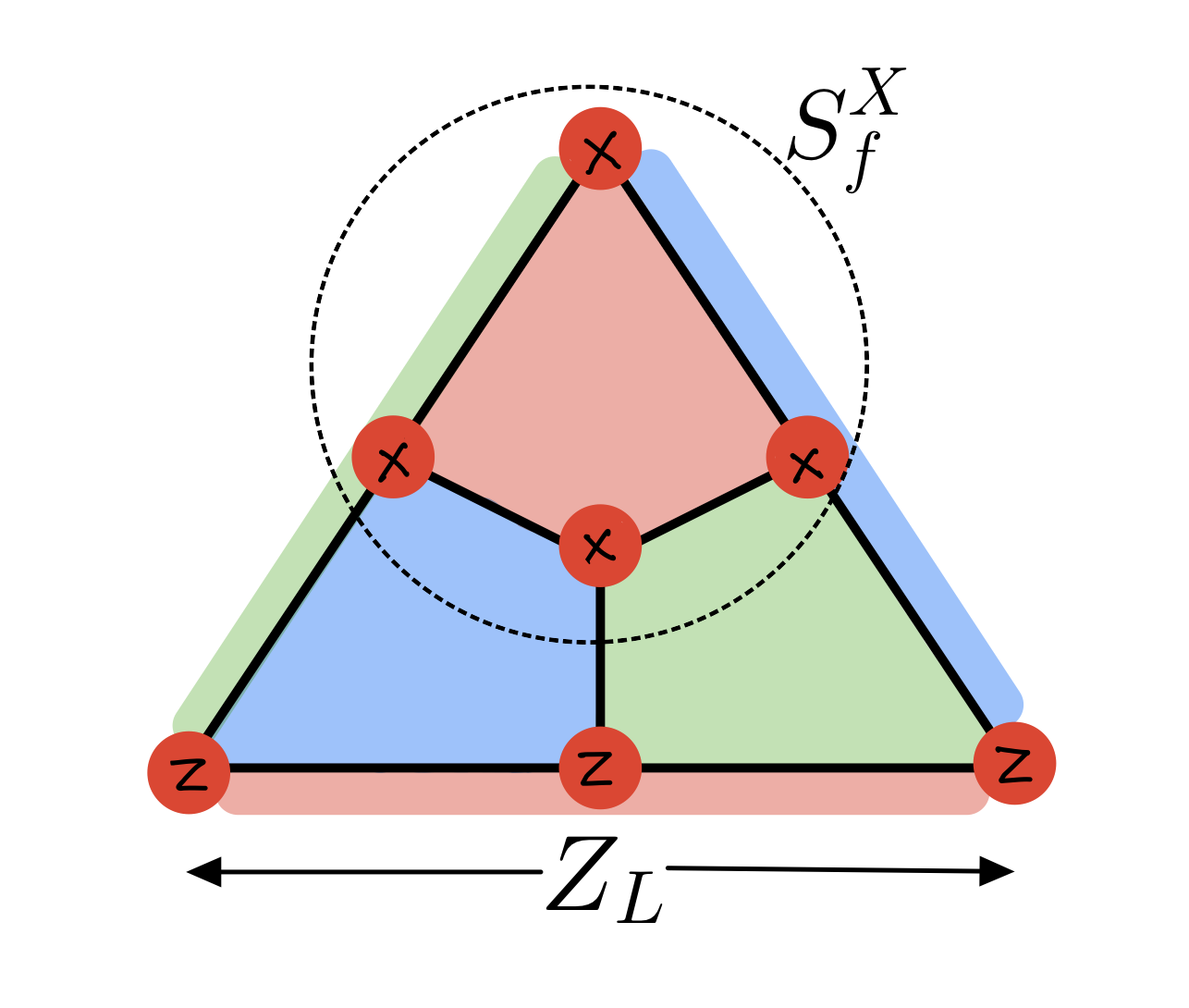}
\caption{The $[[7,1,3]]$ triangular color code. Data qubits (red circles) reside on the vertices, with the three triangular faces shaded red, green, and blue. An example $X$-type stabilizer, $S_f^X$, is shown on the top red face.
The logical $Z$ operator, $Z_L$, is realized as the product of $Z$ operators on the three qubits along any one colored boundary, for example, along the bottom red edge.
}
\label{fig:colorcode}
\end{figure}

The $6.6.6$ color code~\cite{bombin2006topological,fowler2011two,landahl2011fault} is a topological stabilizer code defined on a trivalent lattice in which every face is a hexagon, and the faces can be colored with three colors such that no two adjacent faces share the same color (see Fig.~\ref{fig:colorcode} for the smallest color code of distance $d=3$). 
Qubits are placed on the vertices of the lattice, and each face defines both an \(X\)-type and a \(Z\)-type stabilizer acting on the qubits around that face. This construction enables symmetric protection against bit-flip and phase-flip errors. The $6.6.6$ color code is particularly notable for its high degree of transversal symmetry: all Clifford gates, including the CNOT, Hadamard, and phase (\(S\)) gates, can be implemented transversally, simplifying fault-tolerant gate construction. Logical qubits correspond to nontrivial homological cycles on the lattice, analogous to those in the surface code, but with the advantage that a single layer supports transversal Clifford operations. These properties make the $6.6.6$ color code an attractive candidate for two-dimensional fault-tolerant architectures, as well as for magic-state distillation and code-switching protocols. These advantages motivate us to study how this code can be distributed across multiple QPUs.
In the sub-threshold regime, under concatenated minimum-weight perfect matching (MWPM) decoding~\cite{lee2025color}, the LER decreases exponentially with code distance and is asymptotically expected to scale as
\begin{equation}
 LER(d,p) \propto \left(\frac{p}{p_{\mathrm{th}}}\right)^{d/2},   
\end{equation}
where $p$ denotes the physical error rate and $p_{\mathrm{th}}$ denotes the fault-tolerance threshold of the color code under the chosen decoding scheme.

\subsection{Non-local CNOT through shared EPR pairs}
\begin{figure}[h!]
\centering
  \medskip
  \centering
  \includegraphics[width=0.8\linewidth]{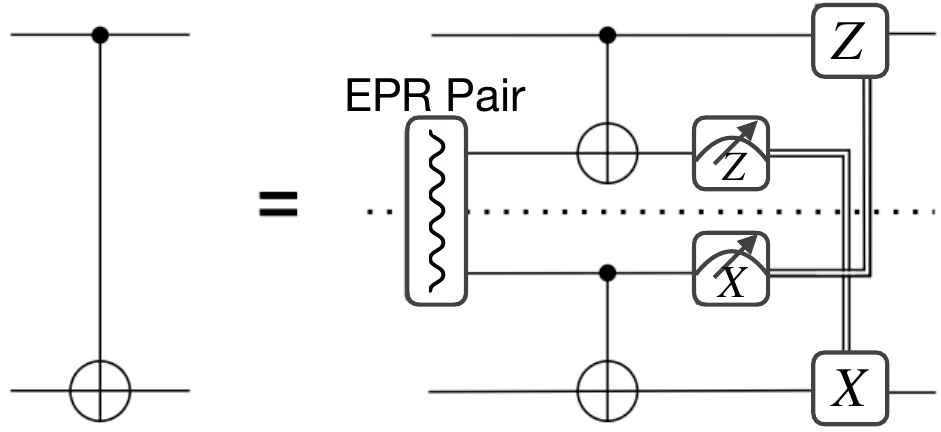}
\caption{Circuit for a remote CNOT. The gate is implemented using a shared EPR pair and local operations, after which the EPR qubits are measured.}
\label{fig:remoteCNOT}
\end{figure}

A non-local CNOT gate allows two distant qubits, each residing in separate quantum modules or QPUs, to interact as if a direct two-qubit gate were available. One standard approach is to use entanglement-assisted gate teleportation~\cite{barral2025review} that employs shared EPR pairs and classical communication. In this protocol, each module holds one half of an EPR pair, and a sequence of local operations and measurements, combined with the exchange of classical bits, implements the logical action of a CNOT between qubits residing in different modules, as shown in Fig.~\ref{fig:remoteCNOT}. This technique effectively transforms spatial separation into a communication overhead rather than a coherence constraint, enabling modular and distributed fault-tolerant architectures. Non-local CNOTs mediated by EPR pairs are fundamental primitives for quantum networks~\cite{bhambay2025optimal,nain2020analysis,vasantam2022throughput}, entanglement-based repeaters~\cite{cacciapuoti2019quantum,caleffi2018quantum}, and DQEC schemes.  
 
\subsection{Heralded entanglement and communication delays}

In distributed quantum architectures, entanglement between remote qubits is typically established via heralded photonic links. Each pair of QPUs repeatedly attempts to generate an entangled Bell pair by emitting photons from local communication qubits and interfering them on a beam splitter. When a specific coincidence pattern of detector clicks occurs, it heralds the successful creation of an entangled pair shared between the two modules. This heralding signal is classical and announces that the Bell pair is ready to be used for non-local operations such as remote CNOTs, without disturbing the entangled state itself. Because photonic emission, transmission, and detection are all probabilistic, the time required to obtain a successful herald varies randomly, leading to stochastic communication latency characterized by the EGR, measured in Bell pairs per second per link.

A low EGR implies long waiting times before a heralded pair becomes available. During these idle periods, data qubits that depend on remote CNOTs must remain coherent while waiting, accumulating idling noise such as dephasing or amplitude damping. Even when heralding succeeds, imperfect photon collection and channel loss can reduce Bell-pair fidelity, introducing additional noise on the seam stabilizers that use them. Therefore, communication delays and imperfect heralding jointly determine the effective reliability and latency of non-local stabilizer measurements. Modeling these effects explicitly is crucial when designing scheduling policies that decide when to measure or skip seam checks, as both waiting and measurement contribute differently to the total error budget in DQEC.

\subsection{Goal: Bridging the gap using efficient scheduling to approach monolithic performance in DQEC}
\begin{figure}[h!]
\centering
  \medskip
  \centering
  \includegraphics[width=11cm]{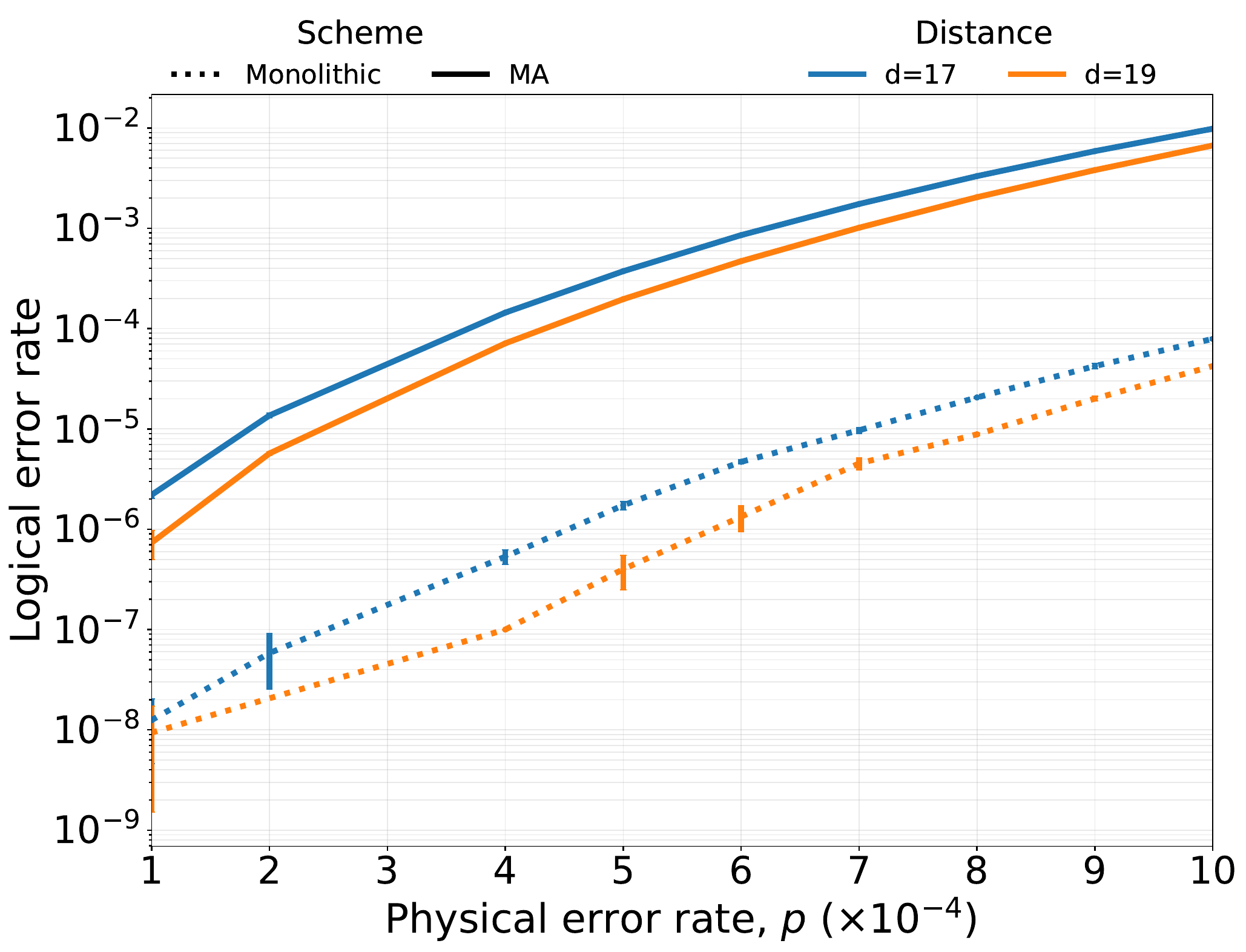}
\caption{
Logical error rate (LER) versus physical error rate for triangular color codes at distances $d\in\{17,19\}$. 
The EGR is fixed at 50 MHz, and $p$ is swept over $[10^{-4},2\times10^{-4},4\times10^{-4},6\times10^{-4},8\times10^{-4},10^{-3}]$.
We compare a monolithic architecture with MA (measure-all in the distributed setting). Increasing the code distance lowers the LER for both schemes. The MA curves lie above the corresponding monolithic curves because non-local seam operations introduce additional noise and waiting-induced decoherence. The y-axis is shown on a logarithmic scale (log spacing).
}
\label{fig:ler_p_d_5_9}
\end{figure}

Fig. \ref{fig:ler_p_d_5_9} shows a clear gap between a monolithic color code and an MA schedule in the distributed setting. At a fixed EGR of $50$ MHz and across the range of physical error rates shown, MA incurs a higher LER because, in every round, it measures all seam checks that require heralded Bell pairs to execute remote CNOTs. The wait to obtain Bell pairs translates into idling errors on data qubits, so MA incurs extra noise each round. Furthermore, the gap grows with code distance because the number of seam operations increases with code distance $d$.

Our goal in this work is to reduce that gap and recover as much of the monolithic performance as possible without requiring a monolithic architecture, by deciding when to measure seam stabilizers using the boundary-aware scheduling schemes SS-$\tau$ and AST.

\begin{figure*}[t!]
  \centering
  \subfloat[\textbf{Distributed QEC setup.} A triangular color code is partitioned across four QPUs connected by photonic links. A central scheduler determines which stabilizers are measured in each round. Bulk checks use only local gates, whereas seam checks span QPU boundaries and require remote entangling operations.\label{fig:1a}]{
    \includegraphics[width=0.35\textwidth]{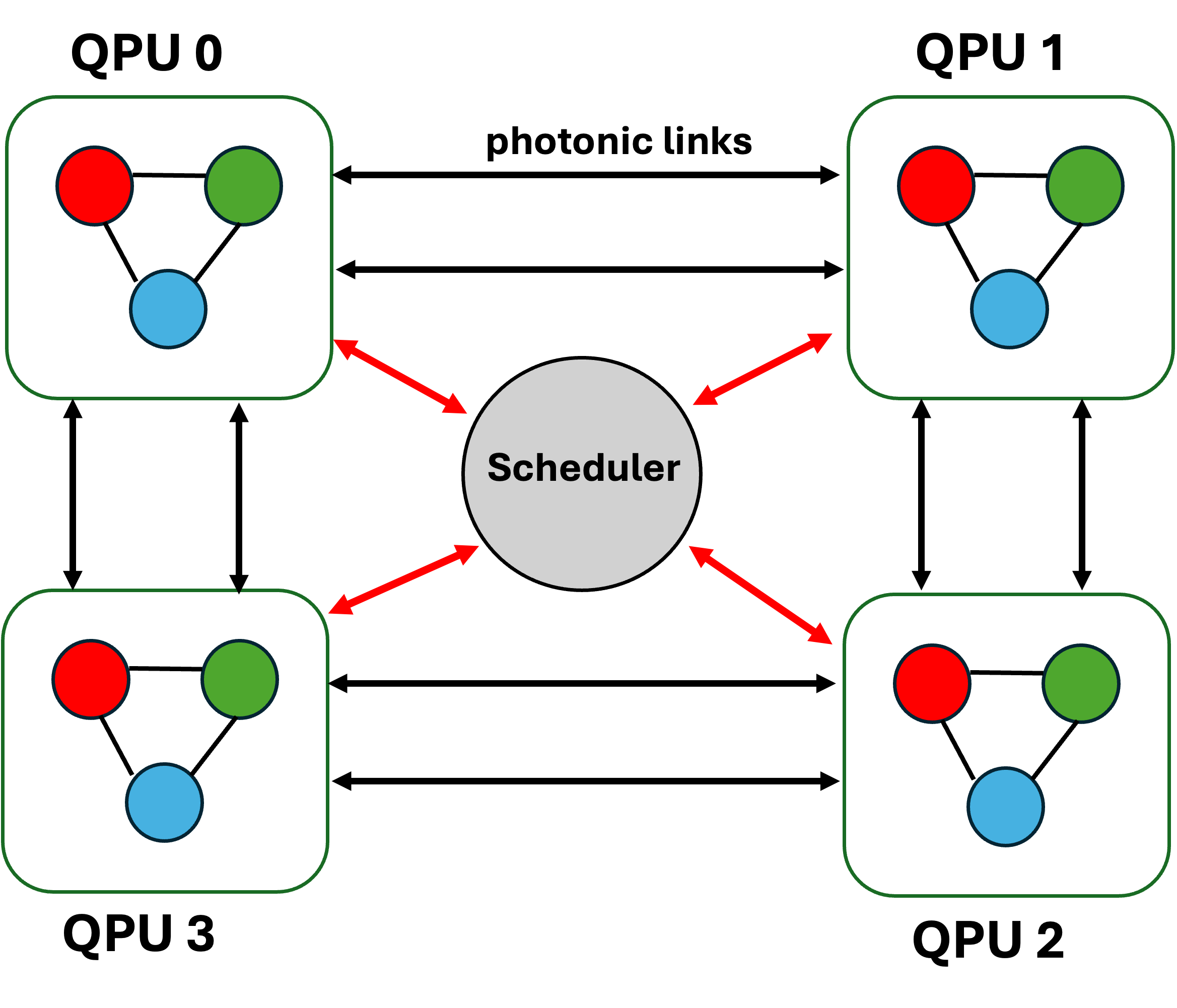}
  }\hfill
  \subfloat[\textbf{Syndrome-measurement schedules.} The timelines compare: (i) \textbf{MA}, in which both bulk checks (BC) and seam checks (SC) are measured every round; and (ii) \textbf{SS-$\tau$}, in which bulk checks are measured every round while seam checks are measured once every $\tau$ rounds. In the intervening rounds, the previous seam outcome is copied forward.\label{fig:abort_shot}]{
    \includegraphics[width=0.6\textwidth]{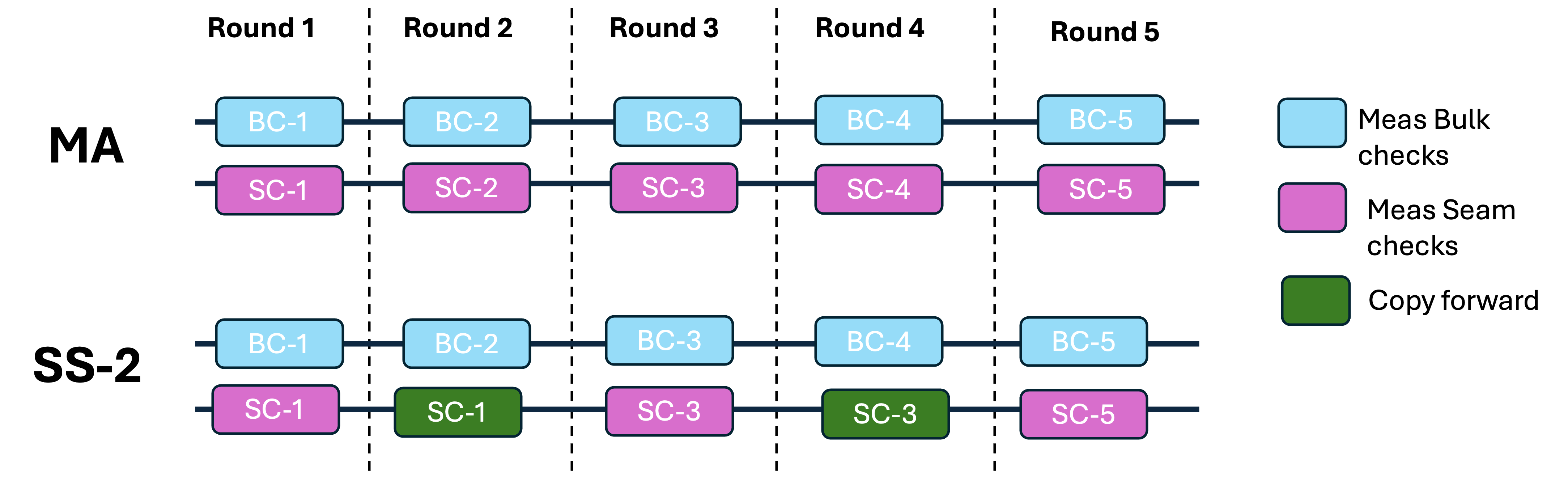}
  }
  \caption{\textbf{Distributed color-code architecture and syndrome-measurement schedules.} (a) A distributed QEC setting with a central scheduler coordinating four QPUs connected by photonic links. Bulk and seam checks arise from locality: bulk checks are local to a single QPU, whereas seam checks cross QPU boundaries. (b) Syndrome-measurement schedules used in this work. BC = bulk check; SC = seam check.}
  \label{fig:dqec_architecture}
\end{figure*}

\section{Distributed QEC Setup and Scheduling Framework}
\label{sec:dqec_frmwrk}

We consider a distance-$d$ triangular color code distributed across four QPUs connected by a photonic interconnect, as shown in Fig.~\ref{fig:dqec_architecture}. Stabilizers are partitioned into \textit{bulk} stabilizers, which lie entirely within a single QPU and are measured using only local CNOTs and readout, and \textit{seam} stabilizers, which span multiple QPUs and therefore require at least one non-local CNOT implemented via a shared Bell pair. This distinction introduces a fundamental asymmetry in the cost of syndrome extraction: bulk checks remain local and are relatively less expensive, whereas seam checks incur additional latency and error due to remote entanglement generation and non-local operations.

QEC proceeds in discrete rounds on a fixed logical clock. In each round, every qubit is subject to physical noise with error rate $p$. When a stabilizer is measured, the corresponding readout has a class-dependent error probability $p_{m,c}$ for $c\in\{\mathrm{bulk},\mathrm{seam}\}$, which incorporates ancilla readout error together with errors arising from the associated entangling gates. For bulk stabilizers these are local CNOT errors, whereas for seam stabilizers they additionally include the effects of remote operations and waiting for Bell-pair generation.

At the centre of the architecture is a scheduling module that determines whether seam stabilizers are measured or skipped in each round. Bulk stabilizers are always measured every round, while seam stabilizers may either be measured or copied forward according to the scheduling policy. When a seam stabilizer is skipped, its ancilla is not entangled and no non-local CNOTs are executed; instead, the most recent available syndrome bit is copied forward. In this way, scheduling changes only the temporal sampling of seam information, while leaving the structure of the decoding problem and, importantly, the decoder itself unchanged. We initialize the system with a measure-all round so that a valid syndrome baseline is available before any skipping occurs.

This design has two important consequences. First, skipping a seam check removes the remote-gate and Bell-pair waiting cost that would otherwise be incurred in that round, thereby reducing interconnect load and avoiding additional idling errors. Second, skipping also lowers the effective sampling rate of seam syndrome information, so parity changes affecting the seam may only be detected after several rounds. Thus, the scheduling problem in distributed QEC is to choose how often seam stabilizers should be refreshed so as to balance remote-operation overhead against syndrome staleness.

The scheduler compiles these decisions into a round plan for both the QPU runtimes and the interconnect controller. If a seam stabilizer is scheduled for measurement, the full stabilizer circuit is executed; if it is skipped, the entangling gates are omitted and the previous syndrome value is reused. In this way, the code and decoder remain unchanged, while the frequency of seam measurements becomes a controllable design parameter.

\subsection{Syndrome measurement schemes}
\label{sec:schemes}
We define three schemes for stabilizer measurements. Seam measurements are performed according to these schemes, and we measure all bulk stabilizers unless otherwise stated.

\subsubsection{Measure-All (MA)}
In every syndrome measurement round, the system measures all stabilizers, both bulk and seam. 
Bulk stabilizers use a local ancilla: prepare the ancilla in the proper basis and apply the six CNOTs in a fixed time-label order, then read the ancilla out to obtain the syndrome bit. Seam stabilizers follow the same logical pattern with some non-local CNOT operations between QPUs via a shared Bell pair. The decoder forms time-like detection events from consecutive outcomes for every face. In this scheme, there is no copying of stabilizers from the previous round, so every stabilizer has fresh information every round. This scheme serves as the baseline for comparing other scheduling schemes.
\subsubsection{Skip-Seam-$\tau$ (SS-$\tau$)}
In this policy, the bulk stabilizers are measured every round as usual, while each seam stabilizer is measured only once every $\tau$ rounds. In the $\tau-1$ measurement rounds, the seam stabiliser values are copied from the most recent actual measurement, so the detector model stays deterministic. Moreover, for skip rounds, we do not entangle the ancilla for that seam check.

By skipping $\tau-1$ rounds of seam stabilizers measurement, we stretch the seam’s time lattice so that consecutive measured outcomes are $\tau$ rounds apart, while bulk faces remain at unit spacing. The inter-QPU resource rate for seams scales down by a factor $1/\tau$, which implies fewer Bell pairs per unit time. 
The detailed SS-$\tau$ scheme for the color code of distance-$d$ is described in Algorithm~\ref{alg:skip-seam-tau}.

It is important to note that choosing $\tau$ balances a trade-off between how often we want to pay the measurement cost of a seam check and how much staleness risk we can accumulate while skipping. Each time we measure a seam stabilizer, it incurs a cost $C_{\text{meas}}$ which includes readout error, remote-CNOT and waiting faults. Spread over $\tau$ rounds, this averages to $C_{\text{meas}}/\tau$ per round. In practice, we must choose $\tau$ that minimises the LER.

\begin{figure}[H]

    \centering
    \begin{subfloat}[Fixed-$\tau$ sweep for $d=11$\label{fig:variable_d_11}]{
        \includegraphics[width=0.8\linewidth]{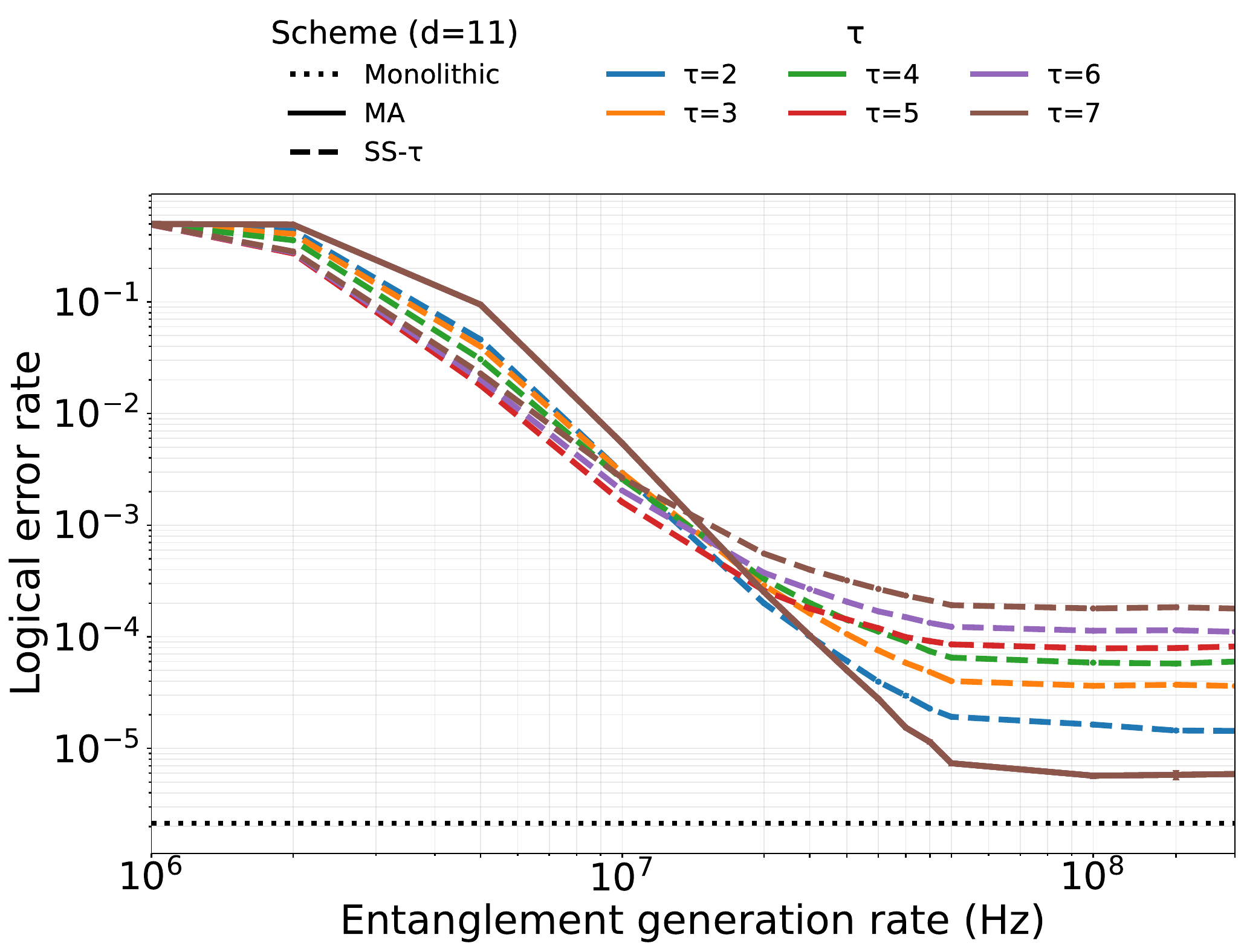}
        }
    \end{subfloat}\hfill

    \centering
    \begin{subfloat}[Fixed-$\tau$ sweep for $d=15$\label{fig:variable_d_15}]{
        \includegraphics[width=0.8\linewidth]{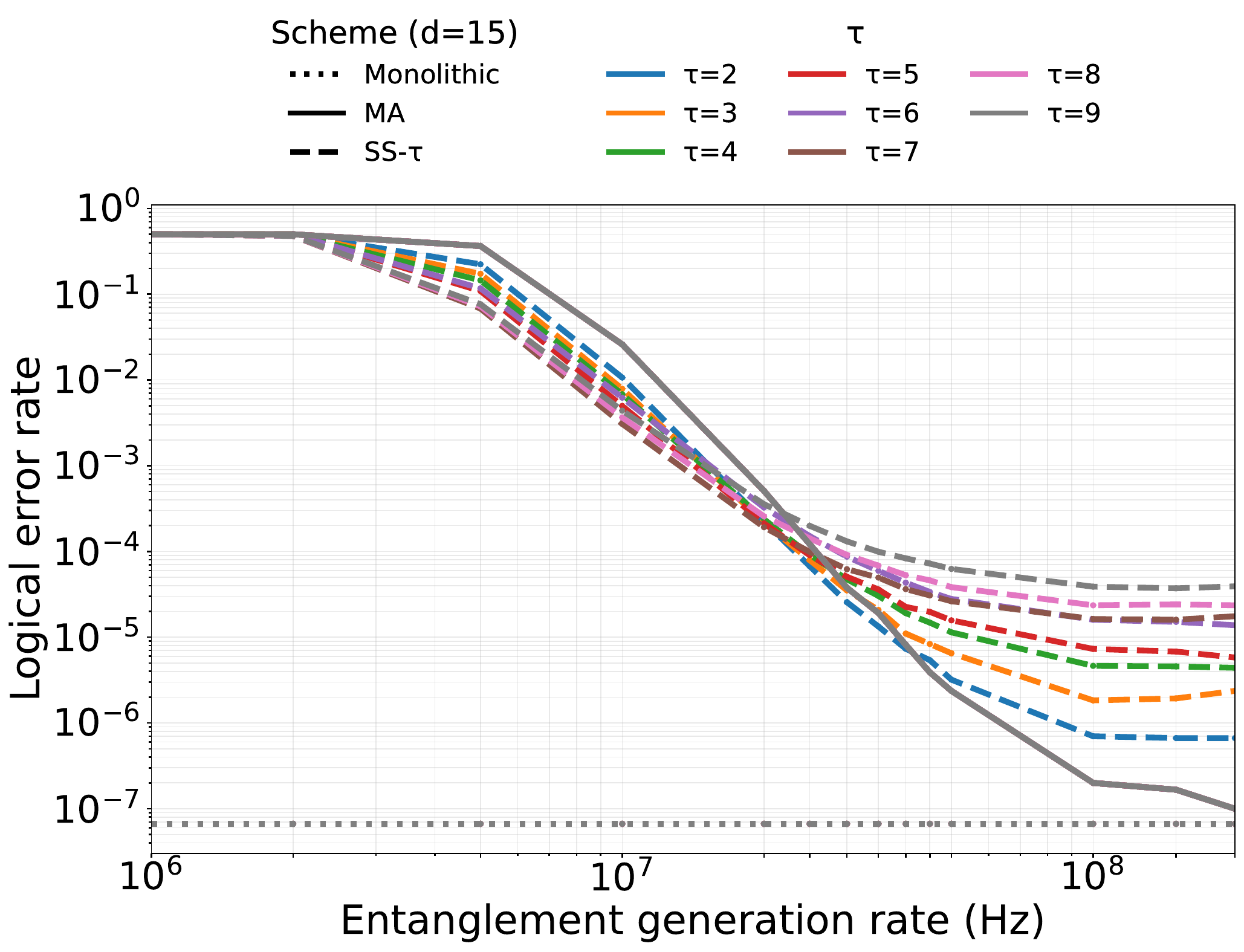}
        }
    \end{subfloat}\hfill

\caption{LER as a function of EGR for fixed seam-refresh intervals $\tau$ at code distances $d\in \{11,15\}$ and $p=10^{-3}$. The optimal seam-refresh interval is not universal, but depends on both code distance and EGR. In the low-EGR regime, larger skip intervals are favoured because they reduce the frequency of costly remote seam measurements and the associated waiting-induced idle noise. In the high-EGR regime, smaller skip intervals are preferred because seam information must be refreshed more frequently to avoid syndrome staleness. Both axes have logarithmic spacing, and EGR is swept over $\{1,2,5,10,20,50,100,150,200\}$ MHz.}
  \label{fig:variable_tau}
\end{figure}

Fig.~\ref{fig:variable_tau} shows that the optimal $\tau$ is not universal: it depends on both the code distance $d$ and EGR. In the low-EGR regime, Bell-pair generation is slow, so the waiting and idling costs of seam measurements are high; in this case, a larger $\tau$ is often preferable because it reduces the frequency of expensive remote operations. By contrast, at medium to high EGR, Bell pairs become available more quickly, the waiting overhead decreases, and smaller values of $\tau$ become favorable since fresher seam information is then more valuable than the savings from skipping. This strong dependence of the best $\tau$ on both $d$ and EGR motivates the design of the AST scheme.


\begin{algorithm}[h!]
\caption{Skip--Seam-$\tau$}
\label{alg:skip-seam-tau}
\begin{algorithmic}[1]
\Require Sets of faces $F_{\mathrm{bulk}}$, $F_{\mathrm{seam}}$ for color code of distnace $d$; period $\tau\geq 1$; number of rounds $T=d$
\Ensure Syndrome stream $(\hat s_{f,t})$ with copy-forward on skipped seam rounds
\Statex \textbf{Notation:}
\Statex \(\hat s_{f,t}\): recorded syndrome bit for face \(f\) at round \(t\).
\Statex \(L_{\mathrm{seam}}\): the most recent round index when the seam was measured.
\State \textbf{Initialize:} Measure all faces in round $t=0$; record $\hat s_{f,0}$ and set $L_{\mathrm{seam}} \gets0$
\For{$t=1,2,\dots,T$}
  \State \textbf{Bulk (Measure every round):}
  \For{$f\in F_{\mathrm{bulk}}$}
    \State   \(\hat s_{f,t} \gets \texttt{Measure}(f)\)
  \EndFor
  \State \textbf{Seam (periodic):}
  \If{$t \bmod \tau = 0$} \Comment{measurement rounds: \(t=\tau,2\tau,\ldots\)}
    \For{$f\in F_{\mathrm{seam}}$}
      \State \(\hat s_{f,t} \gets \texttt{Measure}(f)\)
    \EndFor
    \State \(L_{\mathrm{seam}} \gets t\)
  \Else \Comment{skip entangling/measurement and copy-forward previous true value}
    \For{$f\in F_{\mathrm{seam}}$}
      \State \(\hat s_{f,t} \gets \hat s_{f,L_{\mathrm{seam}}}\)
    \EndFor
  \EndIf
\EndFor
\end{algorithmic}
\end{algorithm}
\subsubsection{Adaptive skip-$\tau$ (AST) scheduling}
\label{ssec:AST}
Under the Adaptive skip-$\tau$ (AST) policy the skip interval depends on the code distance $d$ and the entanglement generation rate $R$. 
The best choice is approximated by two regimes. In the low-EGR regime, remote seam measurements are expensive as it requires heralded Bell-pair generation, and the associated waiting time introduces additional idle decoherence on the data qubits. Since this waiting-induced idling errors are introduced only when the seam is measured, increasing $\tau$ reduces the average per-round remote cost approximately as $1/\tau$. However, taking $\tau$ too large makes the seam syndrome stale. Thus, $\tau$ must balance two competing effects: remote-measurement overhead, which decreases with $\tau$, and syndrome staleness, which increases with $\tau$. In our numerics, we observe that the good candidate for the $\tau^*$ in the low-EGR regime is obtained by refreshing the seam on a timescale comparable to half the code distance, that is, 
\[
\tau^\star \sim \frac{d-1}{2}.
\]
The intuition is that a distance-$d$ code can tolerate only a limited amount of unobserved seam evolution before stale boundary information begins to significantly degrade decoding, so the refresh interval should grow with $d$ but remain well below $d$. The choice $(d-1)/2$ therefore acts as a simple distance-aware rule: larger codes can safely tolerate longer seam gaps, while smaller codes require more frequent refreshes. While this rule provides an effective heuristic in the low-EGR setting, further analysis is needed to determine the best functional forms for these policies.


In the high-EGR regime Bell pairs are generated sufficiently quickly that waiting is no longer the dominant penalty, and the main priority becomes keeping seam information fresh. In that regime, the optimum $\tau^*$ saturates to the small constant $\tau=2$. We therefore define
\begin{equation}
\tau^\star(d,R)=
\begin{cases}
\dfrac{d-1}{2}, & R < R_c, \\[6pt]
2, & R \ge R_c,
\end{cases}
\label{eq:tau_star_piecewise}
\end{equation}
where the crossover entanglement generation rate $R_c$ is defined as the value of $R$ at which
\begin{equation*}
\mathrm{LER}\!\left(d,R,\tau=\frac{d-1}{2}\right)
=
\mathrm{LER}(d,R,\tau=2).
\label{eq:Rc_definition}
\end{equation*}
Equivalently,
\begin{equation*}
R_c(d)
=
\inf\left\{
R:\,
\mathrm{LER}\!\left(d,R,2\right)
\le
\mathrm{LER}\!\left(d,R,\frac{d-1}{2}\right)
\right\}.
\label{eq:Rc_inf_definition}
\end{equation*}
In other words, $R_c$ is the intersection point of the two LER-versus-$R$ curves corresponding to $\tau=(d-1)/2$ and $\tau=2$. 

\begin{algorithm}[t]
\caption{Adaptive skip-seam-$\tau$ scheduling}
\label{alg:adaptive_skip_seam}
\begin{algorithmic}[1]
\Require Code distance $d$, entanglement generation rate $R$, crossover rate $R_c$
\Ensure Adaptive seam-refresh interval $\tau^\star$

\If{$R < R_c$}
    \State $\tau^\star \gets (d-1)/2$
\Else
    \State $\tau^\star \gets 2$
\EndIf

\For{each QEC round $r=1,2,\dots$}
    \If{$r \bmod \tau^\star = 0$}
        \State Measure all seam stabilizers
    \Else
        \State Skip seam stabilizers and measure only local stabilizers
    \EndIf
\EndFor
\end{algorithmic}
\end{algorithm}

\section{Distributing a triangular color code across QPUs}
\label{sec:color_code_part}
In this section, we describe an algorithm for distributing distance-$d$ color-code patches across four QPUs.
We use a balanced partition to ensure that the observed effects are driven by seam-measurement scheduling rather than by load imbalance arising from an uneven distribution of stabilizers among QPUs.

A triangular $6.6.6$ color-code patch of distance $d$ has $n(d)=\frac{3d^2+1}{4}$ data qubits arranged on a triangular lattice.
Our goal is to split these qubits across four QPUs so that (i) most stabilizers remain local to a single QPU, (ii) the inter-QPU seam stabilizers are as short and simple as possible, and (iii) the data-qubit load per QPU is balanced.

To satisfy these three requirements, we first identify a smaller distance-$d'$ color-code patch that fits entirely within one QPU, and then distribute the remaining data qubits across the other three QPUs along the three edges of the larger triangle (see Fig.~\ref{colorcode_partition}). 
Formally, we choose $d'<d$ such that the inner patch contains approximately $N \approx n(d)/4$ data qubits. 
Equivalently, we choose $d'$ such that $d' = \underset{k\in\{3,5,\ldots,d-2\}}{\arg\min} \bigl|n(k) - N\bigr|$.
This allocation makes the load on the QPU that holds the distance-$d'$ inner patch approximately equal to $N$. 
The remaining data qubits are then distributed as evenly as possible across the other three QPUs using a geometric partition of the lattice, specifically a weighted Voronoi method.

 
\begin{figure}[h!]
\centering
  \medskip
  \centering
  \includegraphics[width=9cm]{figures/colorcode_partition.png}
\caption{Balanced four-QPU partition of a $d=9$ triangular color code. Colored dots denote data qubits assigned to each QPU (QPU-$0$ blue, QPU-$1$ red, QPU-$2$ green, QPU-$3$ purple). The dark inner triangle indicates the central subpatch of distance $d'=5$ mapped to QPU-$0$, while the three outer wedge-shaped regions are mapped to other QPUs. 
}
\label{colorcode_partition}
\end{figure}

To place seam ancilla qubits, we use a data-qubit-majority rule. 
For each seam ancilla, we examine its neighboring data qubits. If a majority of those qubits lie on one QPU, we place the ancilla on that QPU; if there is a tie, we break it randomly.
This rule minimizes the number of non-local CNOTs required to execute a seam stabilizer.

\section{Methodology}
\label{sec:methodology}

Our goal in this section is not to provide a hardware-specific threshold estimate, but to isolate and study the trade-off between remote operation overhead and syndrome staleness introduced by seam scheduling. To this end, we adopt a simple circuit-level noise model and a balanced multi-QPU partition. This provides a clear setting to demonstrate the impact of seam measurement scheduling as a proof of concept.

Specifically, we simulate all scheduling policies on distributed triangular color-codes using  Stim with a concatenated MWPM decoder~\cite{lee2025color}. 
For each code distance $d$, we build a circuit-level model of repeated syndrome extraction using $d$ rounds, unless otherwise noted. The color-code is partitioned across multiple QPUs using the method defined in Section~\ref{sec:color_code_part}, where seam stabilizers require remote CNOTs implemented via heralded Bell pairs.
\subsection{Noise Model}
\label{subsec:noise_model}
We adopt a circuit-level Pauli error model~\cite{maurya2025synchronization} that includes wait-time noise on data qubits. 
Concretely, single-qubit gates are followed by a depolarizing channel with probability $p_{1q}$, while two-qubit gates, including remote CNOTs, are followed by a depolarizing channel with probability $p_{2q,c}$ for $c \in \{\mathrm{bulk},\mathrm{seam}\}$. 
Moreover, measurements suffer classical bit-flip errors with probability $p_{m}$. 
In Table \ref{tab:error-model}, we define all gate and readout error probabilities in terms of a single physical error rate $p$.
\begin{table}[t]
\centering
\caption{Error–probability parametrization used in our simulations. All gate and readout errors are tied to a single physical error rate \(p\).}
\label{tab:error-model}
\renewcommand{\arraystretch}{1.2}
\begin{tabular}{ll}
\toprule
\textbf{Symbol} & \textbf{Probability} \\
\midrule
$p_{1q}$ & $p$ \\
$p_{2q,\mathrm{bulk}}$ & $p$ \\
$p_{2q,\mathrm{seam}}$ & $10p$ \\
$p_{m}$ & $p$\\
\bottomrule
\end{tabular}
\end{table}

To capture delays, any idle interval of duration $\Delta t$ is converted, under a Pauli-twirl approximation~\cite{ghosh2012surface}, into a single Pauli channel~\cite{flammia2020efficient} derived from $(T_1,T_2)$. We use 
\begin{align*}
 p_X(\Delta t)&=p_Y(\Delta t)=\tfrac{1}{4}\left(1-e^{-\Delta t/T_2}\right),\\
 p_Z(\Delta t)&=\tfrac{1}{2}\left(1-e^{-\Delta t/T_1}\right)-\tfrac{1}{4}\left(1-e^{-\Delta t/T_2}\right).
\end{align*}
 Idle noise is injected into all the data qubits during network waits for heralded Bell pairs.

\subsection{Hardware timing parameters}
 In Table~\ref{tab:timings}, we list all gate-timing parameters used in our simulations including single and two-qubit gate durations, readout latency, and the per-round cycle time. Unless otherwise noted, these values are taken from an IBM system~\cite{ibm_brisbane_2024}, and are used consistently across all experiments.
 
 \begin{table}[t]
\centering
\caption{System timing parameters used in all simulations.}
\label{tab:timings}
\begin{tabular}{lcc}
\toprule
\textbf{Parameter} & \textbf{Symbol} & \textbf{Value} \\
\midrule
Energy relaxation time & $T_1$ & $200~\mu\text{s}$ \\
Dephasing time & $T_2$ & $150~\mu\text{s}$ \\
Single-qubit gate duration & $T_{1q}$ & $50~\text{ns}$ \\
Two-qubit gate duration & $T_{2q}$ & $70~\text{ns}$ \\
Readout  & $T_{\text{ro}}$ & $1500~\text{ns}$\\
\bottomrule
\end{tabular}
\end{table}
\subsection{Delay to noise calculation}

For a photonic link of length $L$ and fibre loss $\ell$ dB/km, the per-attempt success probability for distributing an EPR pair as a simple attenuation factor $
p_{\text{att}}(L,\ell)
=10^{-\ell(L/1000)}$.
This captures exponential loss with distance in dB units. 

Suppose in a given round we need $n$ remote CNOT pairs. The interconnect has $C$ parallel links, each attempting entanglement at a rate of $EGR$ attempts/second, with a per-attempt success probability of $p_{\text{att}}$. 
Then the expected wait time to create $n$ pairs is
$
\mathbb{E}[W]
\approx
\frac{n^2}{Cp_{\text{att}} EGR}.
$

During the EPR wait, data qubits are not actively part of the round and thus accrue idle time. We model the decoherence contributing to idle time as the longer of the readout latency and the expected EPR wait,
$T_{\text{idle}} = \max\{T_{\mathrm{ro}}, \mathbb{E}[W]\}$.
That is, whichever dominates readout delay or interconnect wait sets the effective idle time.

We convert each idle interval $T_{\text{idle}}$ into a Pauli channel via the Pauli-twirl approximation using $T_1$ and $T_2$ (see Subsection~\ref{subsec:noise_model}).
A standard twirled mapping gives the per-idle probabilities
$p_X(T_{\text{idle}})=p_Y(T_{\text{idle}})$ and 
$p_Z(T_{\text{idle}})$. 
We apply this channel to all data qubits during network waits.

 \subsection{Logical error rate estimation (LER)}
For each configuration, specified by the code distance $d$, scheduling scheme, EGR, and $\tau$, we compile a Stim circuit and perform Monte Carlo sampling. 
Each experiment runs for $S$ shots chosen carefully based on the physical error rate. Syndromes are decoded using a concatenated MWPM decoder, and a shot is counted as a logical failure when the recovered Pauli frame anticommutes with any logical operator of the code. 

\medskip

Overall, the simulation setup follows standard approaches in QEC~\cite{fowler2011two,landahl2011fault} studies, combining a circuit-level noise model, explicit modeling of delays, and repeated syndrome extraction with decoding. The framework captures the key effects relevant to distributed settings, in particular the asymmetry between local and remote operations and the role of waiting-induced decoherence. As such, it provides a suitable setting for demonstrating the impact of seam-measurement scheduling as a proof of concept.

\section{Numerical Results}
\label{sec:num_rslts}


In this section, we evaluate the impact of seam measurement scheduling on LER. We focus on three questions. First, whether scheduling can reduce LER relative to the MA baseline. Second, how this effect depends on EGR. Third, whether scheduling recovers a regime in which increasing code distance improves performance.

We evaluate two seam-scheduling schemes for distributed triangular color codes, SS-$\tau$ and AST, and compare them against the baseline MA policy.
All simulations are performed under the circuit-level noise model described in Section \ref{sec:methodology}, with local depolarising gate noise and additional wait-induced Pauli noise from Bell-pair generation delays on the seam, and using a concatenated MWPM decoder.
We first compare MA and AST as a function of the physical error rate at a fixed EGR.
We then use SS-2 as an example to illustrate why skipping the seam measurement is beneficial in the low-EGR regime, and then evaluate AST across EGR regimes.

\subsection{Logical error rates of scheduling schemes}

\begin{figure}[H]

  \centering
    \includegraphics[width=0.8\linewidth]{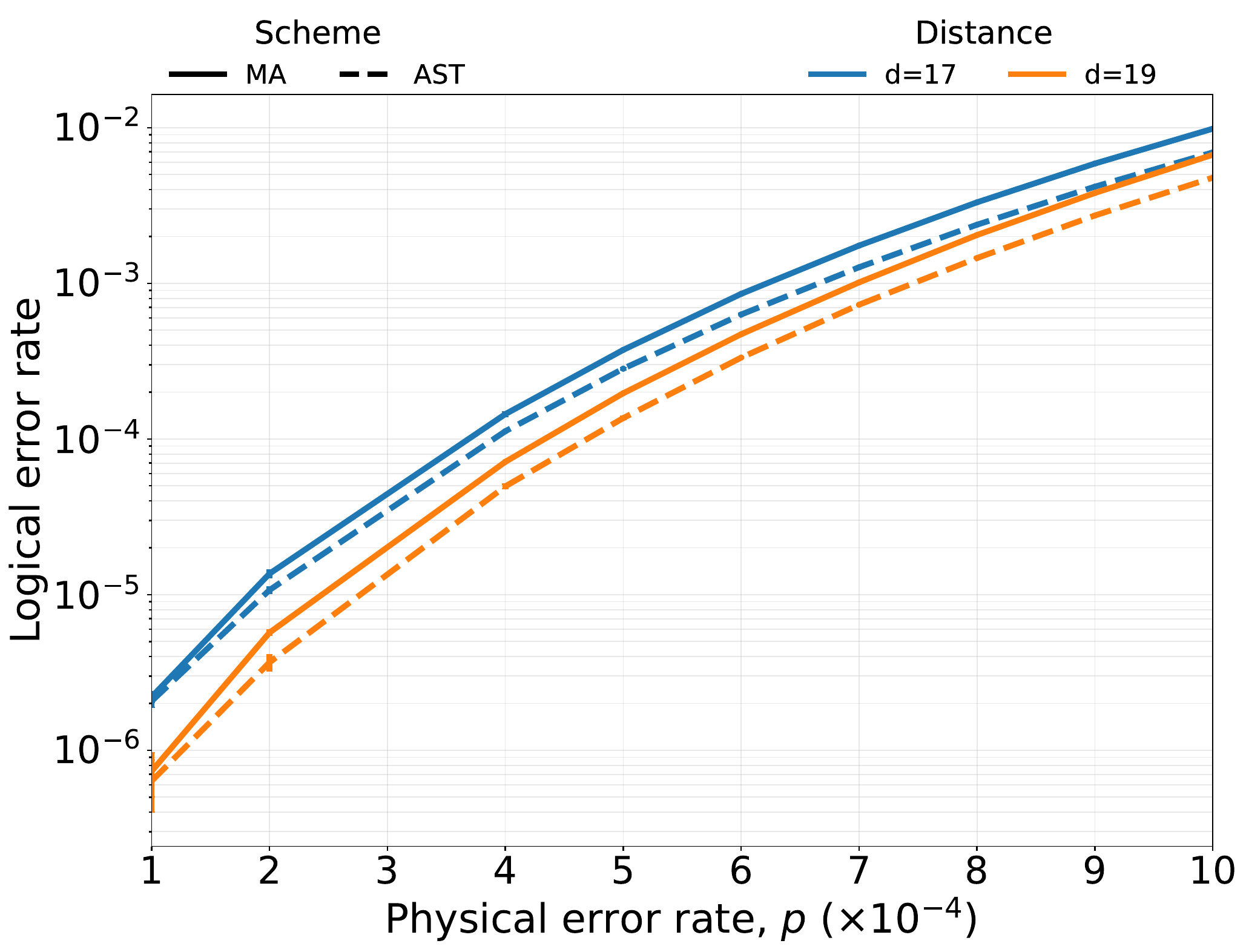}

  \caption{Logical error rate (LER) versus physical error rate for triangular color codes at distances $d\in\{17,19\}$ under the MA and AST scheduling schemes. For both code distances, AST yields a lower LER than MA over the plotted range. The EGR is fixed at 50 MHz, and $p$ is swept over $[10^{-4},2\times10^{-4},4\times10^{-4},6\times10^{-4},8\times10^{-4},10^{-3}]$. The y-axis is shown on a logarithmic scale (log spacing).}
  \label{fig:LERvsP}
\end{figure}

Fig.~\ref{fig:LERvsP}, shows that the AST policy reduces LER relative to the MA baseline at fixed EGR. Here, we compare MA and AST by plotting LER versus physical error rate $p$ for code distances $d\in \{17,19\}$ at fixed EGR = 50 MHz. 
Across the physical error rates considered, AST achieves a lower LER than MA.
This performance arises because AST reduces the frequency of seam measurements in regimes where remote CNOTs and Bell-pair waiting would otherwise introduce substantial additional noise, while still refreshing seam information often enough to maintain decoding performance.
The performance gap between MA and AST is larger at  $d=19$ than at $d=17$.
This suggests that, as code distance increases, the cost of repeatedly measuring seam stabilizers in every round becomes more significant, making adaptive control of the seam-refresh rate increasingly beneficial.

\subsection{LER variability with entanglement generation rate}



\begin{figure}[h!]
    \centering
    \includegraphics[width=0.8\linewidth]{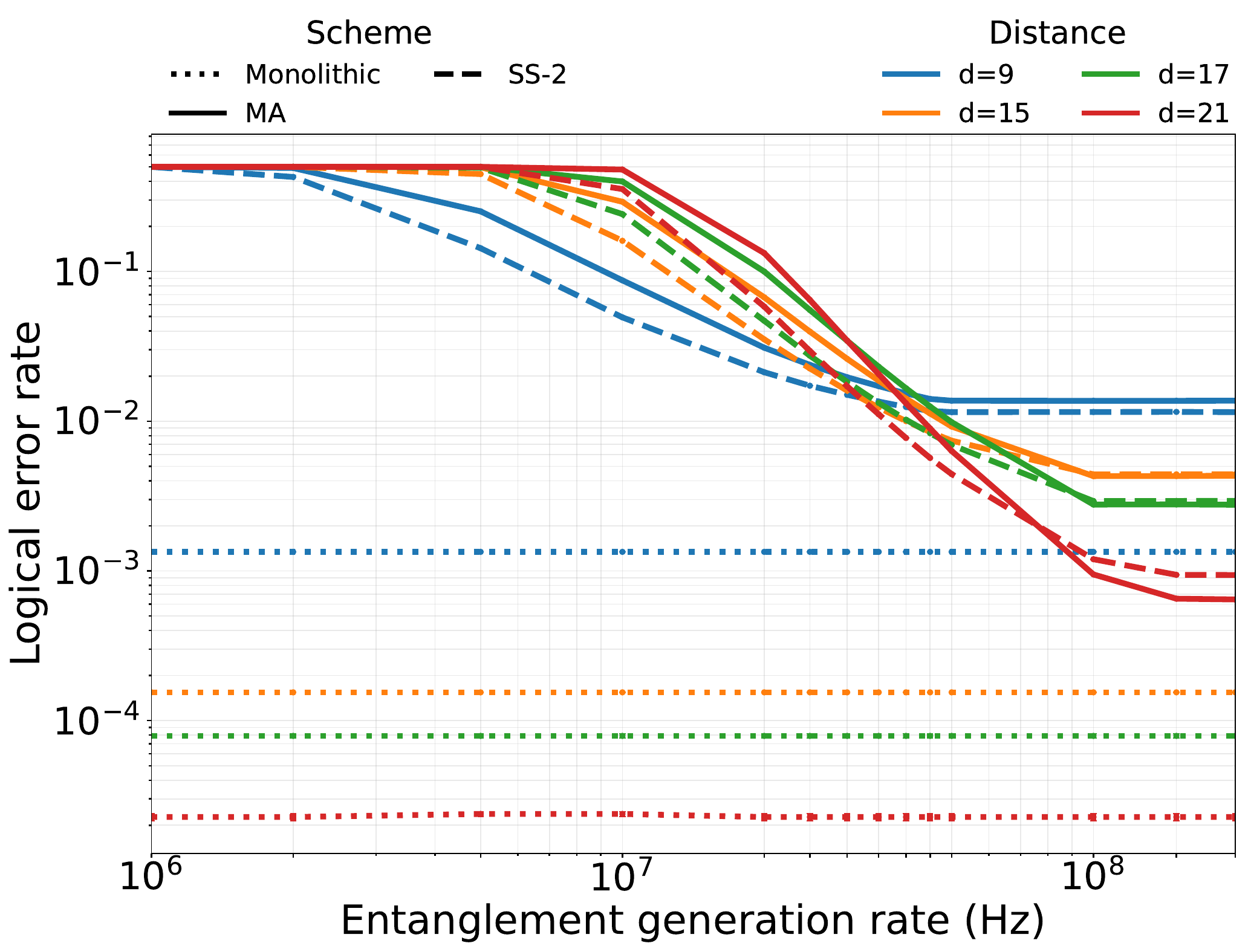}
    \caption{Logical error rate (LER) versus entanglement generation rate (EGR) at fixed $p=10^{-3}$ for triangular color codes at distances $d \in \{9,15,17,21\}$. The monolithic (dotted) baseline is compared with the distributed MA (solid) and SS-$2$ (dashed) schedules. As EGR increases, the LER of the distributed schemes decreases, with SS-$2$ consistently outperforming MA. Both axes have logarithmic spacing, and EGR is swept over $\{1,2,5,10,20,50,100,150,200\}$ MHz.}
    \label{fig:LERvsEGR_p_103_SS2}
\end{figure}
\begin{figure}[h!]
    \centering
    \includegraphics[width=0.8\linewidth]{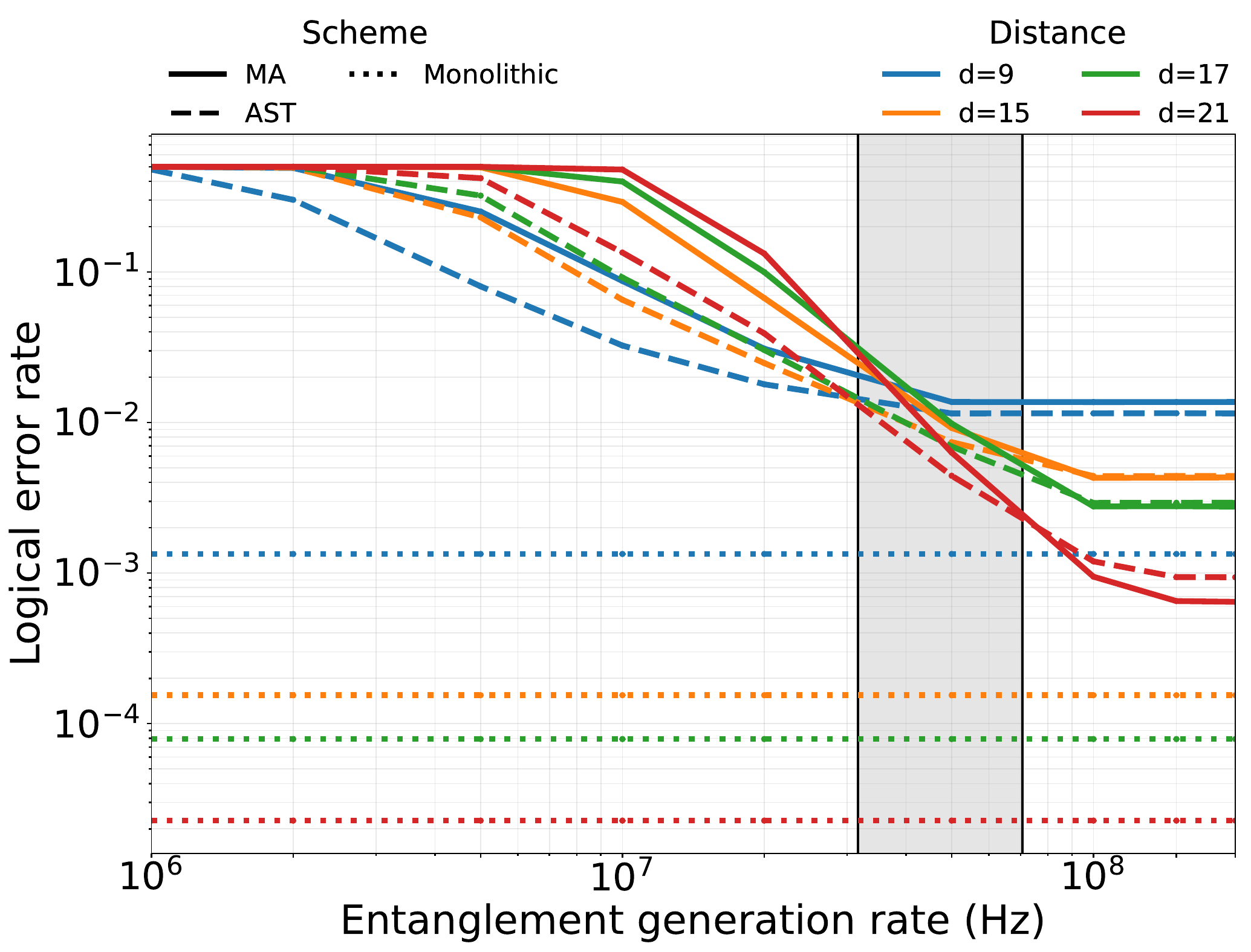}
    \caption{Logical error rate (LER) versus entanglement generation rate (EGR) at fixed $p=10^{-3}$ for triangular color codes at distances $d \in \{9,15,17,21\}$. We compare the monolithic (dotted) baseline with the distributed MA  (solid) and AST (dashed) schedules. Vertical lines and the shaded region indicate the approximate range 31.6--70.7 MHz, within which fault-tolerant distance scaling is observed. Both axes have logarithmic spacing, and EGR is swept over $\{1,2,5,10,20,50,100,150,200\}$ MHz.}
    \label{fig:LERvsEGR_p_103_AST}
\end{figure}

Fig.~\ref{fig:LERvsEGR_p_103_SS2}, provides a motivating fixed-skipping example using SS-2, showing that reducing seam-measurement frequency can improve logical performance when EGR is low.
However, as discussed in Section~\ref{sec:schemes}, this benefit is regime dependent, which motivates the adaptive AST policy considered in Fig.~\ref{fig:LERvsEGR_p_103_AST}.
In Fig.~\ref{fig:LERvsEGR_p_103_AST}, we plot LER as a function of EGR for MA and AST. 
In the low-EGR regime, AST consistently outperforms MA.
When Bell pairs are generated slowly, measuring seam stabilizers in every round incurs substantial waiting time, which translates into additional idle noise on data qubits.
By refreshing the seam less frequently, i.e. skipping with $\tau^{*}=(d-1)/2$, AST reduces this penalty and achieves lower LER.

As EGR increases, Bell-pair generation becomes faster, so waiting-induced idle noise is reduced and the advantage of skipping seam checks diminishes.
In this regime, fresher seam-syndrome information becomes more valuable for decoding, and the gap between AST and MA therefore narrows.
At sufficiently high EGR, the dominant contribution to LER comes less from Bell-pair waiting and more from gate, readout, and residual decoding limitations, so the performance difference between the two scheduling schemes becomes modest.

We also highlight an intermediate EGR window (shaded region) in which the LER decreases as the code distance increases.
To the left of this window, network delays dominate the error budget, and increasing the code distance $d$ actually increases the LER, as the added physical qubits simply accumulate more idling noise.
As the EGR enters the shaded window, this trend reverses, and larger-distance codes achieve lower LER than smaller-distance codes. In this regime, Bell-pair generation is fast enough that waiting-induced idling no longer overwhelms the error budget, while AST still reduces the seam-measurement overhead enough to preserve the advantage of larger code distance.


This behaviour can be understood in terms of two competing effects. At low EGR, waiting induced idle noise dominates the error budget, so increasing the code distance introduces more qubits that accumulate idle noise and therefore increases LER. As EGR increases, this waiting cost is reduced, and the benefit of larger code distance begins to re-emerge. In this intermediate regime, seam scheduling reduces the cost of remote operations while waiting noise is no longer dominant, allowing increasing code distance to again improve error suppression.

\section{Conclusion and Future Directions}
\label{sec:conclusion}




In this work, we studied stabilizer-measurement scheduling for DQEC with triangular color codes distributed across multiple QPUs. In this setting, seam stabilizers require non-local CNOT operations mediated by heralded Bell pairs and are therefore slower and noisier than local bulk stabilizers due to both remote-operation errors and waiting-induced idling noise. As a result, the conventional MA strategy can become suboptimal in distributed regimes where remote entanglement is costly.


To address this, we introduced a boundary-aware scheduling framework that changes how frequently seam stabilizers are measured without modifying either the code or the decoder. Within this framework, the SS-$\tau$ policy shows that even a simple periodic seam-skipping rule can outperform the distributed MA baseline, while the AST policy further demonstrates that adapting the seam-refresh interval to architectural and error-model parameters, such as code distance and EGR, can provide additional gains. Our numerical results show that both SS-$\tau$ and AST reduce LER in regimes where Bell-pair generation is slow and repeated seam measurements incur substantial waiting-induced noise. As entanglement generation becomes faster, this benefit diminishes because fresher seam-syndrome information becomes more valuable for decoding. These results highlight the central trade-off in DQEC between reducing seam-measurement overhead and maintaining sufficiently fresh boundary information.

Overall, our results show that stabilizer measurement scheduling is an important system-level design parameter in distributed QEC. 
More broadly, this work provides a proof of concept that adaptive management of seam-syndrome information can improve  DQEC performance without modifying the underlying code or decoder.
Several directions remain for future work. First, although in this paper we consider simple fixed or piecewise-defined choices of the seam-refresh interval $\tau$, a more complete theory should determine its functional form directly from physical error rates, decoder behavior, and higher-level architectural parameters such as interconnect bandwidth, and modular topology. Second, while we model heterogeneity at the level of bulk versus seam checks, real distributed architectures may exhibit additional non-uniformity arising from the location of the boundary, the physical length and geometry of seam stabilizers. These effects could make different seam checks require different refresh rates rather than a single global $\tau$. Finally, it will be important to study larger modular layouts, more general adaptive scheduling rules, and alternative decoders that can exploit irregularly sampled syndrome information more effectively.


\bibliographystyle{IEEEtran}
\bibliography{refrences} 
\end{document}